\documentclass[slac_one]{revtex4}
\usepackage{graphicx}
\usepackage{fancyhdr}
\begin{document}

\title{Generation of Flavour Dependent Lepton Asymmetries in the E$_6$SSM} 

%

\author{S. F. King}
\affiliation{University of Southampton, Southampton, SO17 1BJ, UK}
\author{R. Luo, D. J. Miller, R. Nevzorov}
\affiliation{University of Glasgow, Glasgow, G12 8QQ, UK}

\begin{abstract}
We calculate flavour dependent lepton asymmetries within
the $E_6$ inspired Supersymmetric Standard Model ($\rm E_6SSM$).
Our analysis reveals that the substantial lepton CP asymmetries
can be induced even if $M_1\simeq 10^6\,\mbox{GeV}$.
\end{abstract}

\maketitle

\thispagestyle{fancy}


\section{INTRODUCTION}
In models with heavy right--handed neutrinos lepton asymmetry can be dynamically generated
via the out--of equilibrium decay of the lightest right--handed neutrino $N_1$. This asymmetry 
subsequently gets converted into the baryon asymmetry due to sphaleron interactions. In the 
standard model and its minimal supersymmetric extension the appropriate amount of baryon asymmetry 
can be induced only if the mass of the lightest right--handed neutrino $M_1\gtrsim 10^9\,\mbox{GeV}$. 
In this case reheat temperature has to be relatively high, i.e. $T_R> 10^9\,\mbox{GeV}$, that results 
in an overproduction of gravitinos. To avoid gravitino problem $T_R$ should be low enough 
$T_R\lesssim 10^{6-7}\,\mbox{GeV}$. Here we study the generation of lepton CP asymmetries within 
the the Exceptional Supersymmetric Standard Model (E$_6$SSM) \cite{e6ssm}.


\section{THE E$_6$SSM}
The E$_6$SSM is based on the $SU(3)_C\times SU(2)_W\times U(1)_Y \times U(1)_N$ 
gauge group which is a subgroup of $E_6$. By definition
$
U(1)_N=\frac{1}{4}\, U(1)_{\chi}+\frac{\sqrt{15}}{4} U(1)_{\psi}\,,
$
where $E_6\to SO(10)\times U(1)_{\psi}$, $SO(10)\to SU(5)\times U(1)_{\chi}$.
To ensure anomaly cancellation the particle content of the E$_6$SSM is extended to
include three complete {\bf $27$} representations of $E_6$ \cite{e6ssm}. The {\bf $27_i$} multiplets 
contain SM family of quarks and leptons, right--handed neutrino $N^c_i$, SM singlet field 
$S_i$ which carry non--zero $U(1)_{N}$ charge, a pair of $SU(2)_W$--doublets $H^d_{i}$ 
and $H^u_{i}$ which have the quantum numbers of Higgs doublets and a pair of colour 
triplets of exotic quarks $\overline{D}_i$ and $D_i$ which can be either diquarks (Model I) 
or leptoquarks (Model II). $H^d_{i}$ and $H^u_{i}$ form either Higgs or inert Higgs multiplets.
In addition to the complete $27_i$ multiplets the low energy particle spectrum of the E$_6$SSM 
is supplemented by $SU(2)_W$ doublet $L_4$ and anti-doublet $\overline{L}_4$ from extra $27'$ 
and $\overline{27'}$ to preserve gauge coupling unification \cite{e6ssm}. 
In the E$_6$SSM the right-handed neutrinos do not participate in the gauge interactions and therefore are 
expected to gain masses at some intermediate scale, while the remaining matter survives 
down to the electroweak scale. The part of the E$_6$SSM superpotential describing the interactions of 
$N_i^c$ with other fields can be written as 
\begin{equation}\label{1}
W_N= h^N_{kxj}(H^u_{k} L_x)N_j^c + g^{N}_{kij}D_{k}d^{c}_{i}N^{c}_{j}\,,
\end{equation}
where $x=1,2,3,4$ while $k,i,j=1,2,3$. In the Model I $g^{N}_{kij}=0$ 
whereas in the Model II all terms in Eq.~(\ref{1}) can be present.

\begin{figure*}[t]
\centering
\includegraphics[width=30mm]{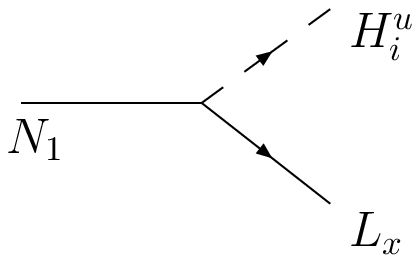}\\
\includegraphics[width=100mm]{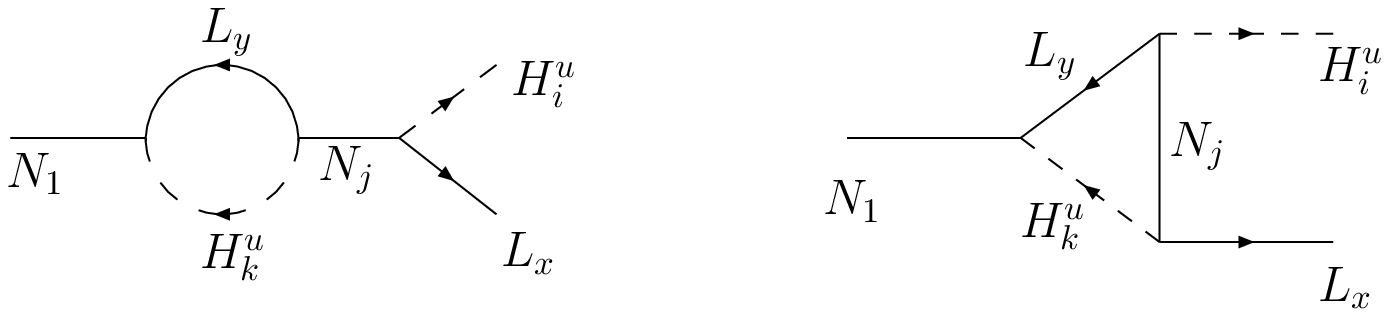}
\includegraphics[width=100mm]{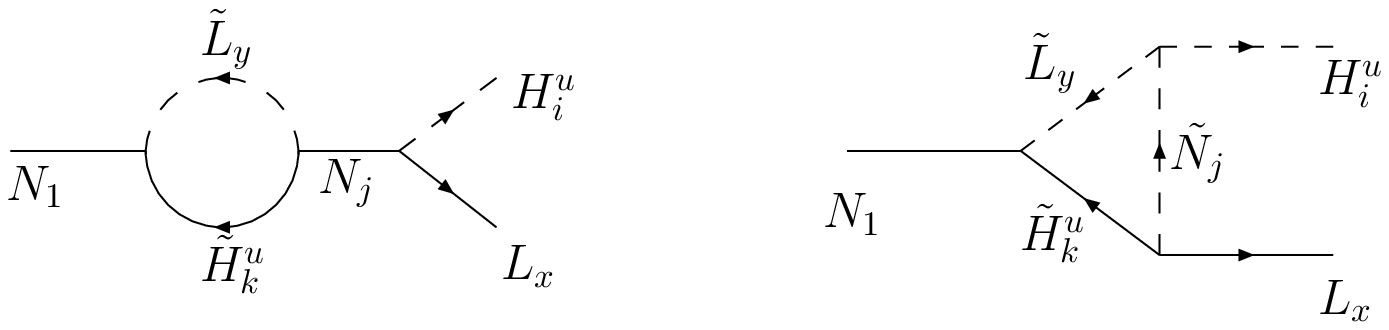}
\includegraphics[width=100mm]{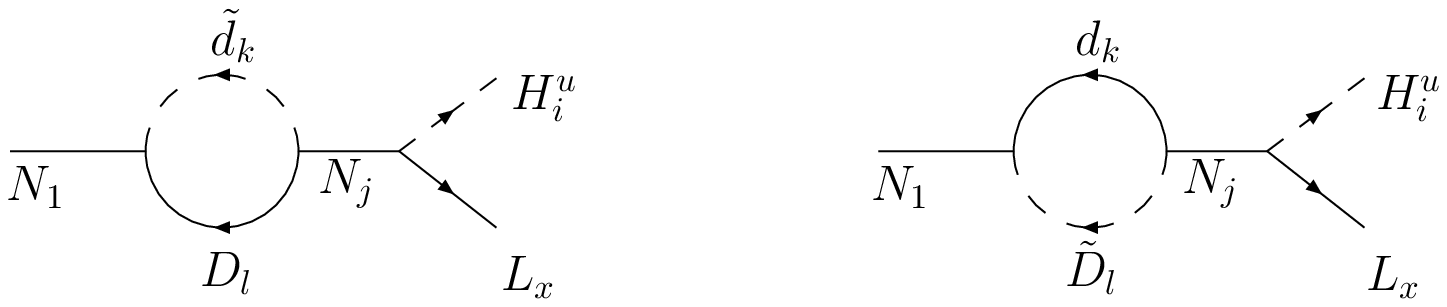}
\caption{Diagrams that give contribution to $\varepsilon^i_{1,\,\ell_k}$.} \label{FeynDiag-1}
\end{figure*}

\section{CP ASYMMETRIES}
The process of the lepton asymmetry generation is controlled by the flavour CP (decay) asymmetries. 
The CP asymmetries associated with the decays of $N_1$ and $\widetilde{N}_1$ into leptons and 
sleptons are given by
\begin{equation}\label{2}
\varepsilon^k_{1,\,f}=\frac{\Gamma^k_{1f}-\Gamma^k_{1\bar{f}}}
{\sum_{m,f'} \left(\Gamma^{m}_{1f'}+\Gamma^{m}_{1\bar{f}'}\right)},\qquad\qquad
\varepsilon^k_{\widetilde{1},\,f}=\frac{\Gamma^k_{\widetilde{1} f}-\Gamma^k_{\widetilde{1} \bar{f}}}
{\sum_{m,\,f'} \left(\Gamma^m_{\widetilde{1} f'}+\Gamma^m_{\widetilde{1} \bar{f}'}\right)},
\end{equation}
where $f$ and $f'$ correspond to either lepton or slepton field while $\Gamma^k_{1f}$ and 
$\Gamma^k_{\widetilde{1}f}$ are partial decay widths of $N_1\to f +H^u_k$ and $\widetilde{N}_1\to f+H^{u}_k$.
At the tree level CP asymmetries vanish. The non--zero contributions to the CP asymmetries arise from the 
interference between the tree--level amplitudes of the lightest right--handed neutrino decays and 
one--loop corrections to them. Supersymmetry ensures that
$
\varepsilon^k_{1,\,\ell_k}=\varepsilon^k_{1,\,\widetilde{\ell}_k}=\varepsilon^k_{\widetilde{1},\,\ell_k}=
\varepsilon^k_{\widetilde{1},\,\widetilde{\ell}_k}\,.
$
The tree--level and one--loop diagrams that contribute to the $\varepsilon^k_{1,\,\ell_k}$ are shown in Fig.~1.
When $M_1\ll M_2, M_3$ we get (see \cite{King:2008qb})
\begin{equation}\label{4}
\varepsilon^{k}_{1,\,\ell_x}=-\frac{1}{8\pi A_1}\sum_{j=2,3}\mbox{Im}\biggl[2 A_j h^{N*}_{kx1} h^{N}_{kxj}+
\sum_{m,\,y} h^{N*}_{my1} h^{N}_{mxj}  h^{N}_{kyj} h^{N*}_{kx1}\biggr]
\left(\frac{M_1}{M_j}\right)\,,
\end{equation}
where $A_j=\sum_{m,y} h^{N*}_{my1} h^{N}_{myj} + \frac{3}{2}\sum_{m,n} g^{N*}_{mn1} g^{N}_{mnj}$
if $j=2,3$ while $A_1=\sum_{m,y} h^{N*}_{my1} h^{N}_{my1}$.

\begin{figure*}[t]
\centering
\includegraphics[width=30mm]{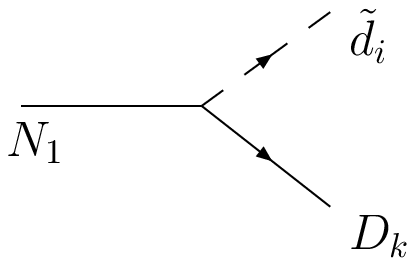}\\
\includegraphics[width=100mm]{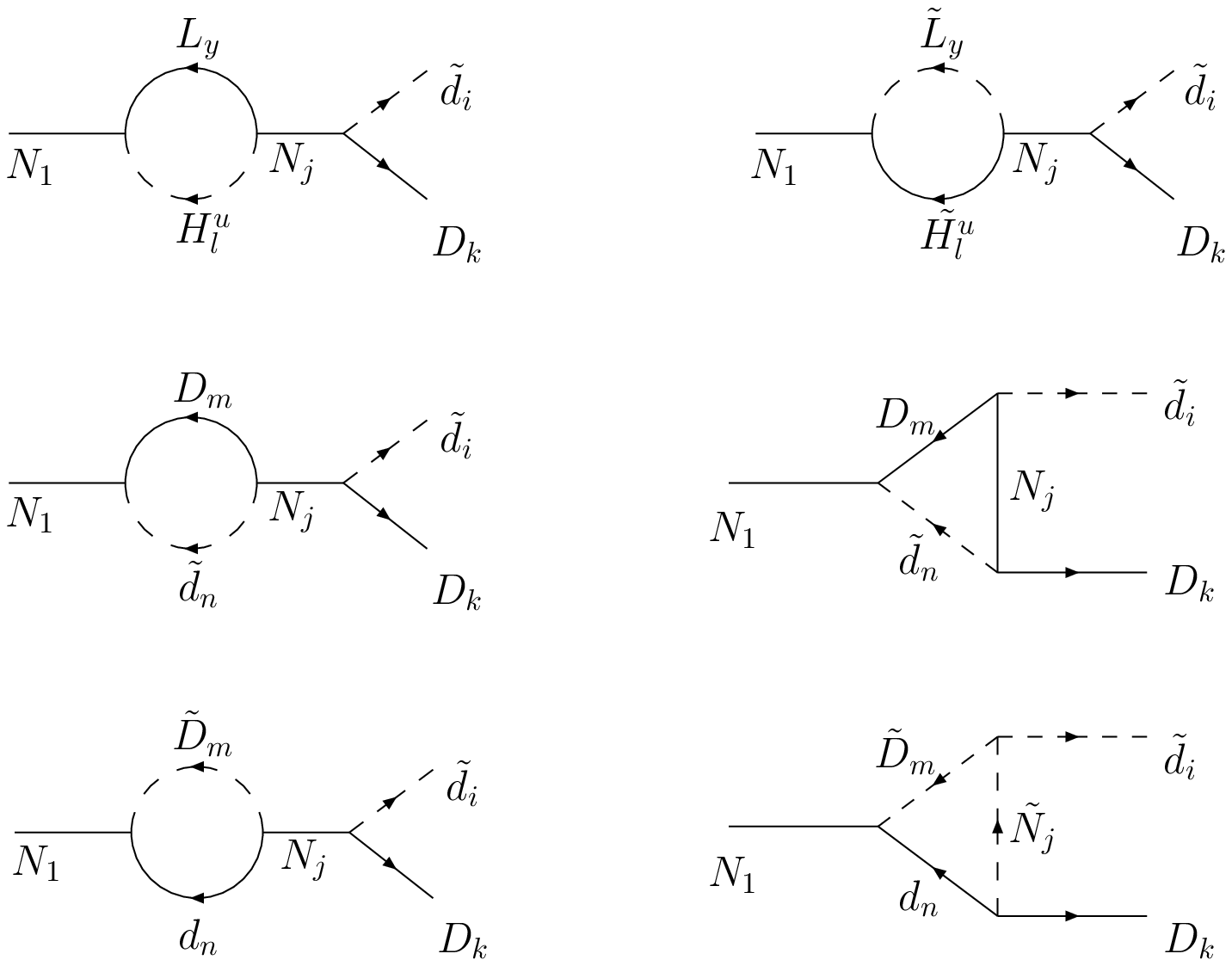}\\
\caption{Diagrams that give contribution to $\varepsilon^{i}_{1,\,D_k}$.} \label{FeynDiag-2}
\end{figure*}

The CP asymmetries associated with the decays of $N_1$ and $\widetilde{N}_1$ into exotic 
quarks (squarks) can be defined similarly to the lepton ones. 
The interference of the tree--level decay amplitude with the one--loop corrections (see Fig.~2) 
yields
\begin{equation}\label{5}
\varepsilon^{i}_{1,\,D_k}=-\frac{1}{8\pi A_0}\sum_{j=2,3}\mbox{Im}\biggl[2 A_j g^{N}_{kij} g^{N*}_{ki1}+
\sum_{m,\,n} g^{N*}_{mn1} g^{N}_{mij} g^{N}_{knj} g^{N*}_{ki1}\biggr]\left(\frac{M_1}{M_j}\right)\,,
\end{equation}
where $A_0=\sum_{k,\,i}g^{N}_{ki1} g^{N*}_{ki1}$.

\section{RESULTS AND CONCLUSIONS}
To simplify our analysis we impose $Z^H_2$ symmetry under which all superfields except 
$H_d\equiv H^d_{3}$, $H_u\equiv H^u_{3}$ and $S\equiv S_3$ are odd. The $Z^H_2$ symmetry 
allows to avoid unacceptably large non-diagonal flavour transitions and reduces 
the part of 
the E$_6$SSM superpotential that describes the interactions of $N_i^c$ with other superfields
\begin{equation}\label{6}
W_N\to h^N_{3xj}(H_{u} L_x)N_j^c.
\end{equation}
We also assume that the right--handed neutrino mass scale is relatively low ($M_1\sim 10^6\,\mbox{GeV}$), 
so that $h^N_{3ij}$ are negligibly small ($i=1,2,3$), and $M_3\gg M_1, M_2$. In this approximation 
$\varepsilon^{3}_{1,\,L_4}$ is much larger than other CP asymmetries. In the considered case
the maximal absolute value of $\varepsilon^{3}_{1,\,L_4}$ is given by 
\begin{equation}\label{7}
|\varepsilon^{k}_{1,\,\ell_4}|= \frac{3 M_1}{8\pi M_2} |h^{N}_{342}|^2,
\end{equation}
where $h^{N}_{342}$ is a coupling of the second lightest right--handed neutrino to $H_u$ and $L_4$.
From Eq.~(\ref{7}) one can see that in the considered case the substantial lepton decay asymmetry 
$\varepsilon^{k}_{1,\,\ell_4}$ ($\sim 10^{-6}-10^{-4}$) can be induced even for $M_1\simeq 10^6\,\mbox{GeV}$ 
if $|h^{N}_{342}|$ vary from $0.01$ to $0.1$. At low energies the asymmetry generated in the $L_4$ 
sector gets converted into the ordinary lepton asymmetries via the decays of $L_4$. 
This suggests that in the E$_6$SSM successful thermal leptogenesis can be achieved without 
encountering the gravitino problem.

\begin{acknowledgments}
We would like to thank S. Antusch, C. D. Froggatt, L. B. Okun and D.~Sutherland for fruitful discussions.

RN acknowledges support from the SHEFC grant HR03020 SUPA 36878.
\end{acknowledgments}

\end{document}